%% file: quark02.tex
\documentclass[12pt]{article}
\usepackage[english]{babel}
\usepackage[figuresright]{rotating}
\usepackage{epsfig}
\usepackage[intlimits,tbtags]{amsmath}
\usepackage{amssymb,amscd}
\begin{document}
\include{quark02-tit}
\include{quark02-txt}

%
%
\end{document}

%% file: quark02-tit.tex
%
%
\begin{titlepage}
\thispagestyle{empty}

\title{
Physics Results and Future Plans \\of \\the ZEUS Experiment
\footnote{Presented at the 12$^{th}$ International Seminar on High Energy Physics, Novgorod the Great, Russia, June 1 -- 7, 2002 }
}                                                       
                    
\author{Janusz Chwastowski\\[2mm]
        Henryk Niewodnicza\'nski Institute of Nuclear Physics,\\
        {\it ul. Radzikowskiego 152,}\\
        {\it 31 - 342 Krak\'ow, Poland}\\[4mm]
        {\it for The ZEUS Collaboration}\\
}
\date{3.\ June 2002}

\maketitle
\begin{abstract}
Selected ZEUS results on the proton and the photon structures are presented.
A short outline of the future measurements, the HERA machine and the ZEUS 
detector status is given.
\end{abstract}
\end{titlepage}

%% file: quark02-txt.tex
\pagenumbering{arabic}
\pagestyle{plain}
\section{Introduction}
\label{intro}
During the first decade HERA was predominantly run with 27.5 GeV leptons 
against 820 GeV protons. From 1998 on, the proton beam energy was 
increased to 920 GeV. During this time HERA reached the record value 
of the instantaneous luminosity
${\cal L} = 2\cdot 10^{31} ~{\rm sec}^{-1}{\rm cm}^{-2}$.
The ZEUS detector \cite{zeusdet} collected 110 $pb^{-1}$ of the integrated 
luminosity out of which about 16 $pb^{-1}$ were taken with the electron beam.\\
For the neutral current (NC) reaction, $ep \rightarrow e'+X$, the four-momentum
transfer can be calculated as $Q^{2} = -q^2 =  -(k-k')^{2} $ where $k$ and $k'$
are the incident and the scattered lepton four-momentum, respectively. The 
fraction $x$ of the proton momentum carried by the struck  quark is 
$x = {Q^{2}}/({2P\cdot q})$ with $P$ denoting the proton four-momentum. The 
inelasticity is $y = ({q\cdot P})/({k\cdot P})$ and $W^{2} = (q+P)^{2}$ 
measures the energy of the hadronic system.\\
The HERA experiments extended the kinematic range by more than two orders of 
magnitude in both, $x$ and $Q^{2}$ with respect to that accessible to the 
earlier, fixed target experiments.
\section{Proton Structure}
\label{protstruc}
The neutral current cross section can be written in terms of the structure 
functions $F_2$, $F_{L}$ and $xF_{3}$ as
$$\frac{d\sigma^{e^{\pm}p}}{dxdQ^{2}} = \frac{2\pi\alpha^{2}}{xQ^{4}}
\left[Y_{+}F_2(x,Q^2)-y^{2}F_{L}(x,Q^2)\mp Y_{-}xF_{3}(x,Q^2)
\right]$$
where $Y_{\pm} = 1\pm(1-y)^{2}$.
For $Q^{2}$ values much below the $Z^{0}$ mass the parity violating $xF_{3}$ is
negligible. In LO QCD $F_2$ can be interpreted as a sum of the quark 
contributions weighted with the quark charge squared. For large $Q^{2}$ and not
to small $x$ the contribution of the longitudinal function $F_{L}$ is small.\\ 
\begin{figure}[ht]
\vspace{-1cm} 
{\epsfig{file=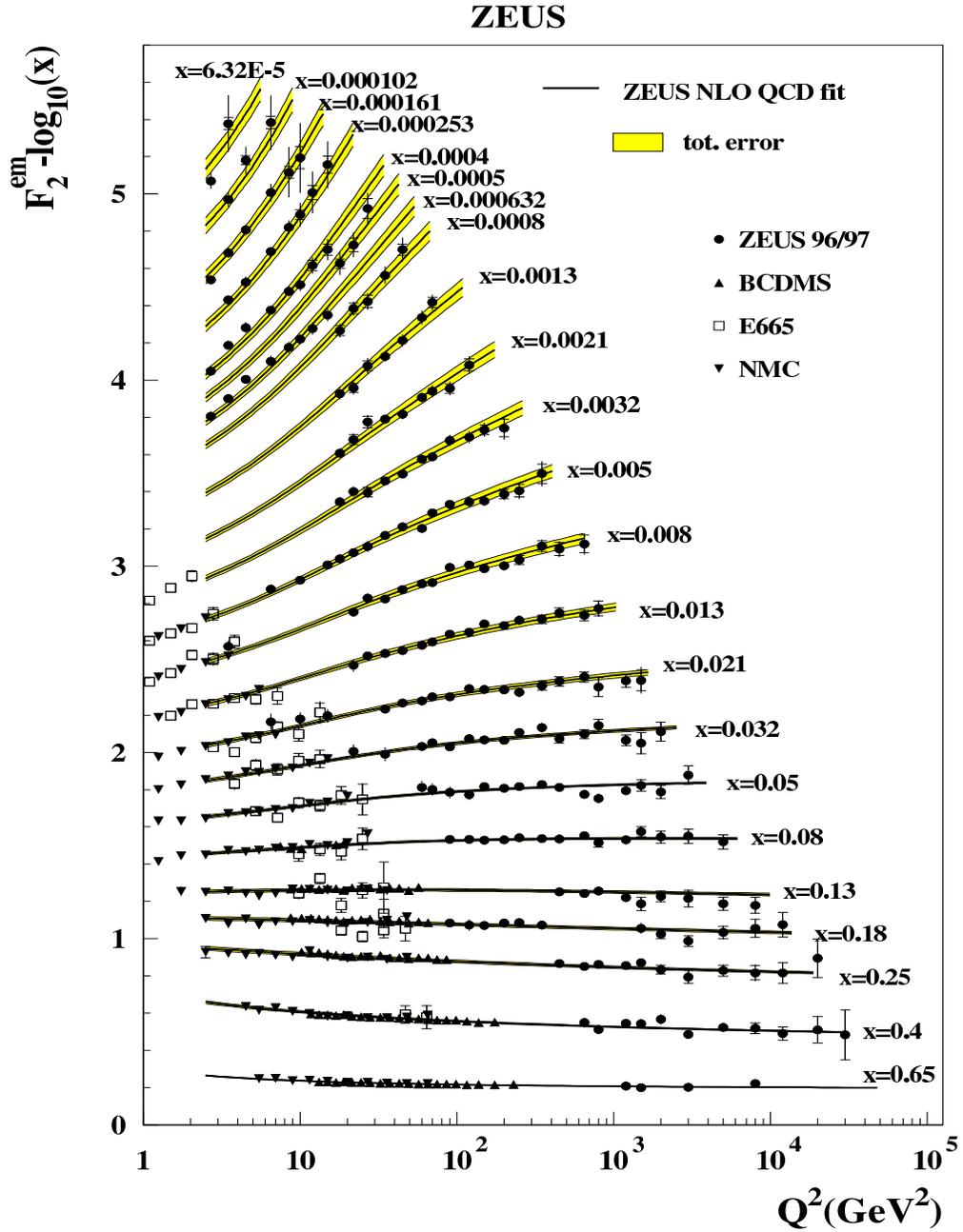,width=14cm,height=18cm}}
\caption{The ZEUS NLO QCD fit compared to the ZEUS 96/97 and fixed-target 
$F_2$ data.}
\label{fig:f2}
\end{figure}
\begin{figure}[hb]
{\epsfig{file=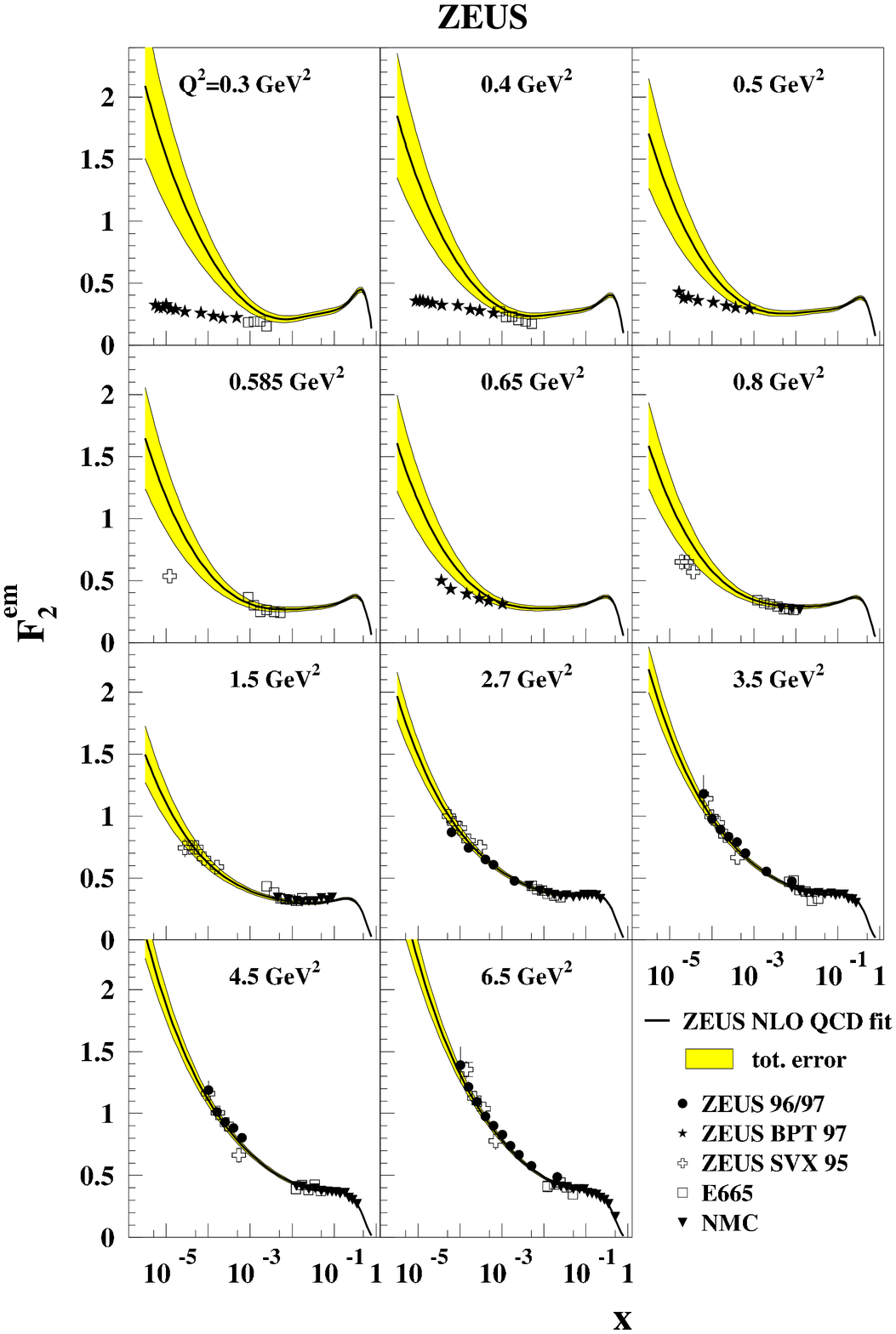,width=14cm,height=18cm}}
\caption{The ZEUS NLO QCD fit compared to the ZEUS high precision low 
$Q^{2}$ data.}
\label{fig:svtx}
\end{figure}
\indent
The ZEUS high precision data on $F_{2}$ are presented in Fig. \ref{fig:f2}
together with the fixed target experiments data. One can observe that $F_{2}$ 
slightly decreases with $Q^{2}$ for high $x$ reflecting $q \rightarrow qg$
transitions. It becomes approximately $Q^{2}$ independent for $x \approx 0.2$
and steeply increases with $Q^{2}$ for decreasing $x$. This strong scaling 
violation reflects the influence of the $g \rightarrow qq$ splitting.\\
These data were used in the NLO DGLAP analysis  \cite{tassi} to determine the
parton distributions.  The analysis was performed for 
$6.5\cdot10^{-5} < x < 0.65$,
\clearpage 
\noindent
$2.5 < Q^{2} < 30000$ GeV$^2$ and to diminish the higher 
twist contribution $W^{2} > 20$ GeV was required. The evolution was carried out
in terms of the singlet, non - singlet and gluon distributions. Fixed - target 
data were used to constrain the fits at high $x$ and provide information on the
valence distributions and the flavor composition of the sea.

The data are very well described by the fits. It was checked  that the ZEUS NLO
parametrisations of the parton distributions are in good  agreement with other
\cite{cteq,mrst} parametrisations. The gluon distribution was found to be
valence like at small $Q^{2}$ and for increasing $Q^{2}$ it shows very strong
increase with decreasing $x$. The sea distribution is flat at small $Q^{2}$ and
it increases with decreasing $x$ for higher values of $Q^{2}$. However, this
increase is much weaker than the one seen in case of the gluon distribution.
The longitudinal structure function $F_{L}$ is valence like for relatively
small $Q^{2}$. At $Q^{2} \approx 1$ GeV$^{2}$ it appears to be negative however
compatible with zero within the errors. The fit results were compared to high
precision low $Q^{2}$ data \cite{bpcdata}. This comparison is shown in Fig.
\ref{fig:svtx}. The data are well reproduced for $Q^{2} \gtrsim 1.5$ GeV$^{2}$.
Below, the discrepancies between the data and the fit results are large. Also, 
$F_{L}$ shows a non-physical behaviour for small $Q^{2}$. This clearly shows 
limitations of the DGLAP fit application.
\begin{figure}[h]
\epsfysize=7.5cm
\epsfxsize=7.5cm
\centerline{\epsffile{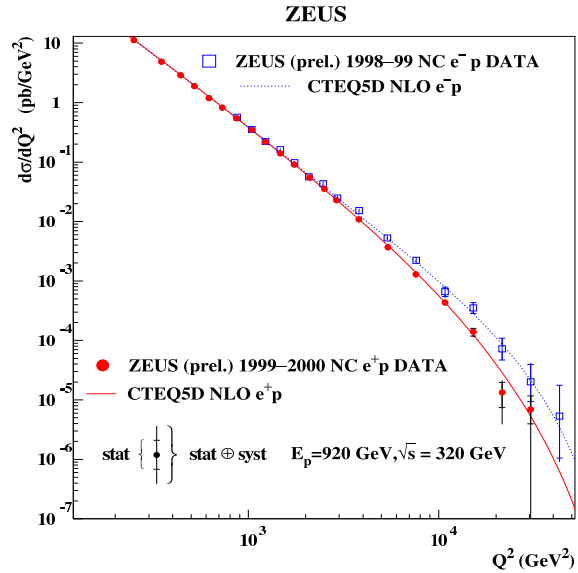}}
\caption{The differential $e^{\pm}p$ NC cross sections compared to the SM
expectations evaluated using CTEQ5D PDF.}
\label{fig:ncxs}
\end{figure}

The neutral current cross section was measured for $Q^{2} > 185$ GeV$^2$
\cite{moritz}. The data are presented in Fig. \ref{fig:ncxs} together with the 
Standard Model (SM) predictions. A clear dominance of the photon exchange is 
seen for $Q^{2} \leq 3000$ GeV$^{2}$. Above this value the $\gamma - Z^{0}$
interference becomes visible. The data are well described by the the
parametrisations and the ZEUS NLO fit results. However, particularly, for high
$Q^{2}$ they are statistically limited. The $xF_3$ structure function was
extracted for $Q^{2} > 3000$ GeV$^{2}$ and was found to be well described by
the SM motivated parametrisations.
\begin{figure}[h]
\epsfysize=12cm
\epsfxsize=10cm
\centerline{\epsffile{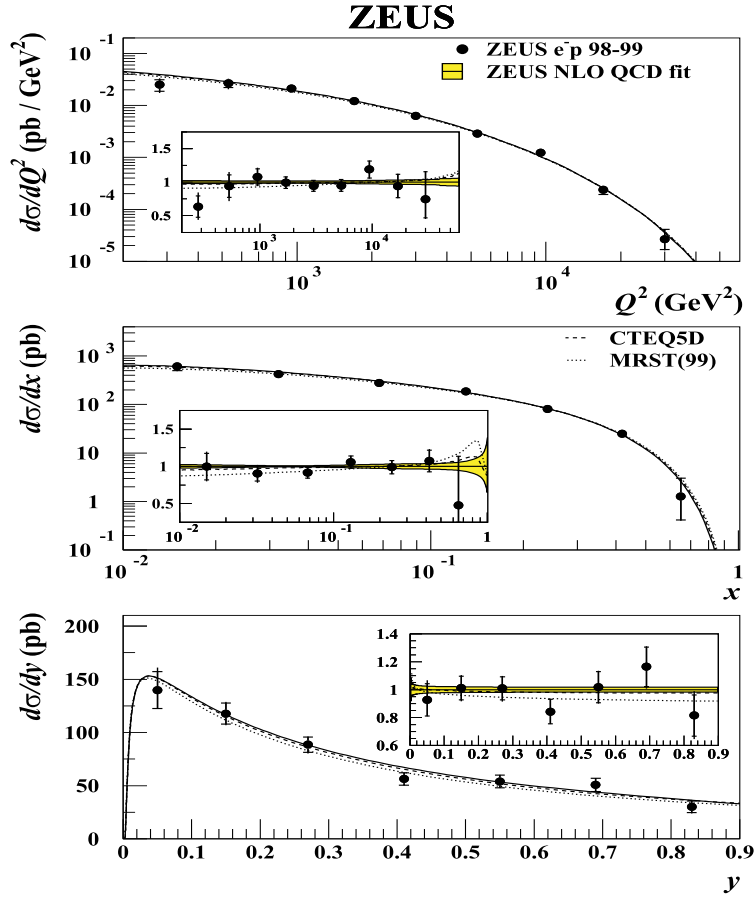}}
\caption{The $e^-p$ CC cross sections compared to the ZEUS NLO QCD fit, 
CTEQ5D and MRST(99) PDFs. The insets show the ratios of the measured cross 
sections to the SM expectations evaluated using ZEUS NLO QCD fit.} 
\label{fig:ccxs}
\end{figure}

The cross section for the charged current (CC) reaction, 
$ep \rightarrow \nu +X$, was measured for $Q^{2} > 200$ GeV$^{2}$ \cite{sjors}.
The data together with the ZEUS NLO and other parametrisations are shown in 
Fig. \ref{fig:ccxs}. The data are very well described by the parametrisations 
over several orders of magnitude. They were used to extract value of the mass 
of the $W$ boson -- $M_{W}$. The fit was performed to the shape of the cross 
section. In the fit the Fermi constant was fixed to its $PDG$ value 
$G_{F} = 1.11639 \cdot 10^{-5}$ GeV$^{-2}$ \cite{pdg}. The fit yielded 
$M_{W}=80.3\pm 2.1(stat.)\pm 1.2(syst.)\pm 1.0(pdf)$ GeV for the electron data.This value well compares to the positron data result of: 
$M_{W}=81.4^{+2.7}_{-2.6}(stat.)\pm 2.0(syst.)^{+3.3}_{-3.0}(pdf)$ GeV 
\cite{posmw}, the $PDG$ value: $M_{W} = 80.422 \pm 0.047$ GeV and to the H1 
result \cite{h1wm}.
\section{Photon Structure}
\label{photstruc}
In photoproduction at HERA a quasi -- real photon emitted by the electron 
interacts with a proton. The LO QCD divides such interactions into two classes.
In resolved processes a photon acts as source of partons and only a fraction of
its momentum, $x_\gamma$, participates in hard scattering. In direct processes
the photon interacts via the boson -- gluon fusion or the QCD Compton and acts
as a point-like particle with $x_\gamma \approx 1$. The measurements of the jet
cross section in photoproduction \cite{dijetpap,h1dj} are sensitive to the 
proton and photon structures, and to the dynamics of the hard sub-process.

\begin{figure}[ht]
\epsfysize=11cm
\epsfxsize=11cm
\centerline{\epsffile{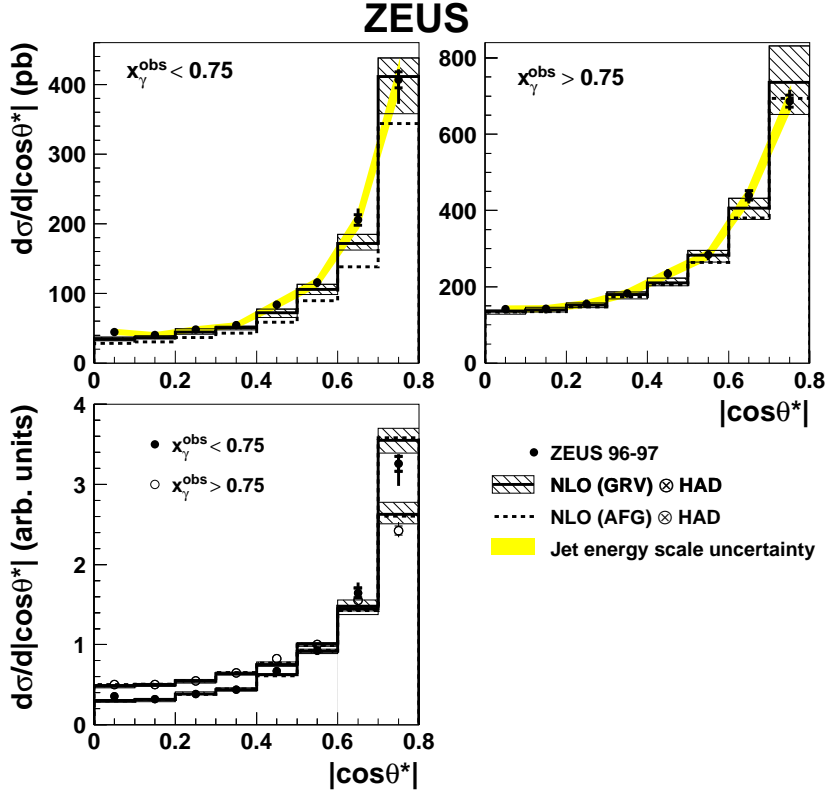}}
\caption{The $| \cos{\theta^{\ast}}|$ distribution for jets for the resolved ($x^{obs}_\gamma < 0.75$) and direct ($x^{obs}_\gamma > 0.75$) components dominated regions, and comparison of the shapes of the $| \cos{\theta^{\ast}}|$ 
distributions (left bottom).}
\label{fig:costh}
\end{figure}
\noindent
Since $x_\gamma$ is not measured directly then $x^{obs}_\gamma$ - the fraction
of the photon momentum participating in the production of the two highest 
energy jets was introduced \cite{xobs}
$$ 
x^{obs}_\gamma =
\frac{E^{jet1}_T e^{-\eta_{jet1}}+E^{jet2}_T e^{-\eta_{jet2}}}
     {2~y~E_e}
$$
where $E^{jet1,2}_T$ are the transverse energies of the jets in the LAB frame,
$\eta_{jet1,2}$ are the jets pseudorapidities and $y$ is the fraction of the
lepton energy, $E_e$, carried by photon in the proton rest frame. In the LO QCD
$x^{obs}_\gamma =  x_\gamma$. It was shown \cite{dijetpap} that the resolved
component dominated region for $x^{obs}_\gamma < 0.75$. Above this value the
direct component dominates.\\
The distribution of angle, $\Theta^\ast$, between the jets in the parton -- 
parton CMS is sensitive to the form of the matrix element. 
For direct processes, mediated by the quark, the distribution
$d\sigma/d\cos{\Theta^\ast} \sim (1-cos{\Theta^\ast})^{-2}$. If the process is
mediated by the gluon exchange, like in case of the resolved component, then
$d\sigma/d\cos{\Theta^\ast} \sim (1-cos{\Theta^\ast})^{-1}$. 
Fig. \ref{fig:costh} shows the angular distributions in the resolved and direct
component dominated regions together with the NLO QCD predictions of
\cite{frixione} which include different parametrisations of the photon PDFs:
GRV-HO \cite{grv} and APG-HO \cite{apg}. The CTEQ5M1 \cite{cteq5m1} proton PDF
was used in the calculations and the hadronisation corrections were estimated 
with HERWIG \cite{herwig} and PYTHIA \cite{pythia} Monte Carlos.\\
For $x_\gamma^{obs} < 0.75$ the data are underestimated by the calculations'
results by about 10-20\%. The GRV-HO delivers better description of the data.
For $x_\gamma^{obs} > 0.75$ the data are well reproduced by the calculations.
The shapes of the angular distributions for both regions are compared in Fig.
\ref{fig:costh} (left bottom). Also the results of the NLO calculations are 
shown. In general the data are well reproduced by the calculations. The 
distribution measured for $x_\gamma^{obs} < 0.75$ is steeper than the one for
$x_\gamma^{obs} > 0.75$. This confirms the differences in the dominant 
propagators.
\begin{figure}[ht]
\centerline{
\epsfysize=7.5cm
\epsfxsize=6.5cm
\epsffile{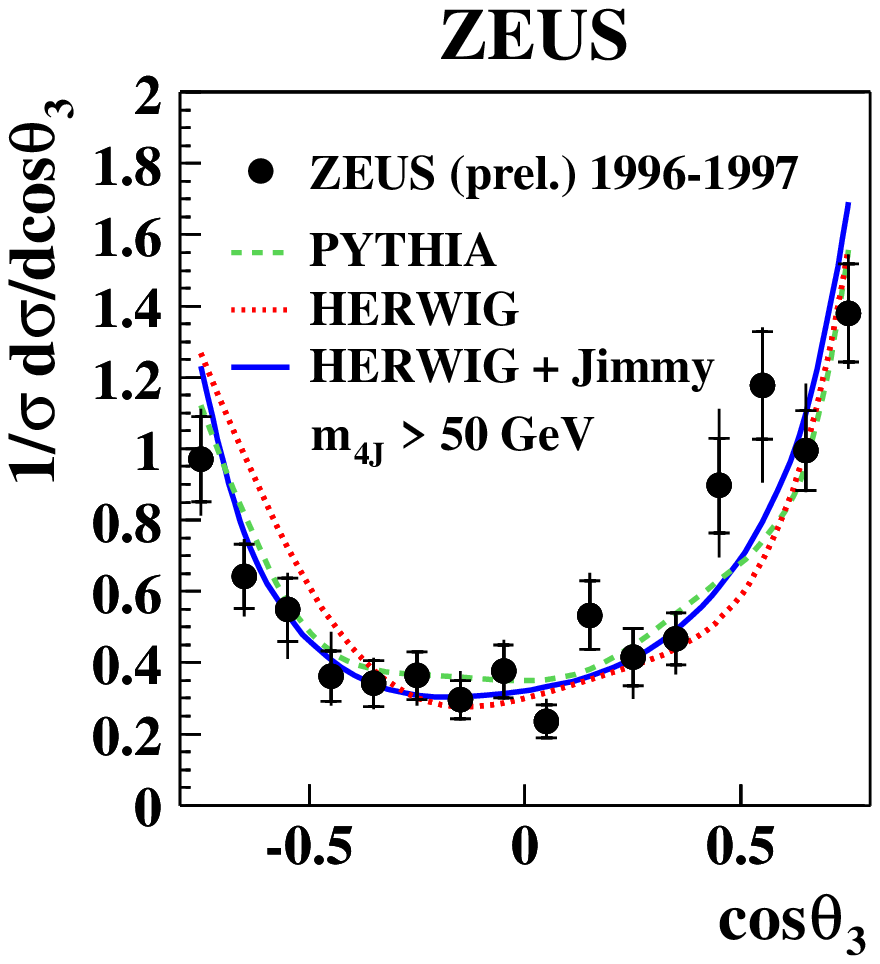}
\hspace{0.5cm}
\epsfysize=7.5cm
\epsfxsize=6.5cm
{\epsffile{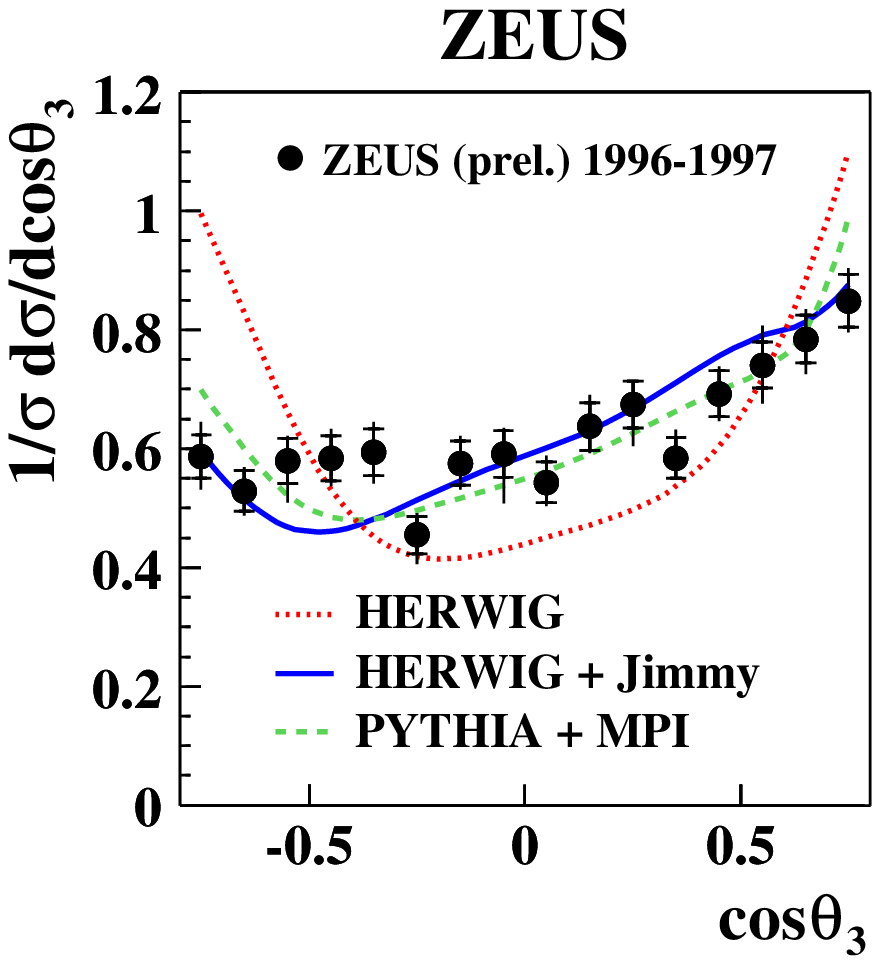}}
}
\caption{The $\cos{\theta_{3}}$ distribution for the high jets mass region ($M_{4j} > 50$ GeV)
(left) and for ``inclusive'' sample (right).}
\label{fig:cost3m}
\end{figure}

The photon structure can also be looked at in multi-jet events. The four 
jet production is sensitive to QCD at $\alpha_s^3$ and to the multi-parton
interactions in the final state. Since it is difficult to manipulate four
objects then the number of jets is reduced. Namely, two jets with the lowest
invariant mass are combined into one. Afterward, the jets are re-ordered in
energy. The dynamics is investigated looking at the distribution of
$cos{\Theta_3}$, the angle between the proton and the most energetic jet
directions. This distribution is shown in Fig. \ref{fig:cost3m} together with
the MC models' predictions.In the ``perturbative'' region, for the four-jet 
mass $M_{4j} > 50$ GeV, the distribution is peaked in the forward and backward
directions. The data are well described by the shown MC irrespectively whether
or not the multi-parton interaction (MPI) option is included into the 
calculations. In case of the ``inclusive'' sample the distribution increases 
with increasing $cos{\Theta_3}$. The data are well described by the MC models 
with the MPI option included. The HERWIG without the MPI fails to reproduce the
data.

The total cross section for the photoproduction was recently re-measured
\cite{sigtot}. The data used came from dedicated runs. This allowed the control
and reduction of the systematic errors. The total cross section measured at 
$W = 209$ GeV was found to be
$$\sigma_{ToT}^{\gamma~p} = 174 \pm 1(stat.)\pm13(syst.) \mu{\rm b}.$$
The result is compatible with the H1 result \cite{h1sigtot} and the value
predicted by the Donnachie and Landshoff \cite{dl}.
\section{Future Plans}
\label{future}
The increase of the luminosity and the longitudinal polarisation for ZEUS and 
H1 are the main goals of the HERA upgrade programme. These goals are achieved 
by the introduction of the mini beta focusing scheme (about 3.5 times increase
in luminosity) and the spin rotators. It is expected to collect 1000 pb$^{-1}$ 
during few years of running. The luminosity will be shared equally between the 
different lepton beam charges and polarisations. Runs with decreased proton 
beam energy for the $F_{L}$ and high $x$ measurements are foreseen. Presently, 
the main stress is put on the background reduction in the experimental areas 
and the beams' currents increase.\\
\noindent
The physics programme was discussed in great detail in \cite{proceed}. Below
only few examples are given. In general the precision tests require 3-10 times
increase in the luminosity. Longitudinal polarisation of the lepton beam  will
help to pin down some problems.\\
\begin{figure}[h]
\epsfysize=7cm
\epsfxsize=9cm
\center{\epsffile{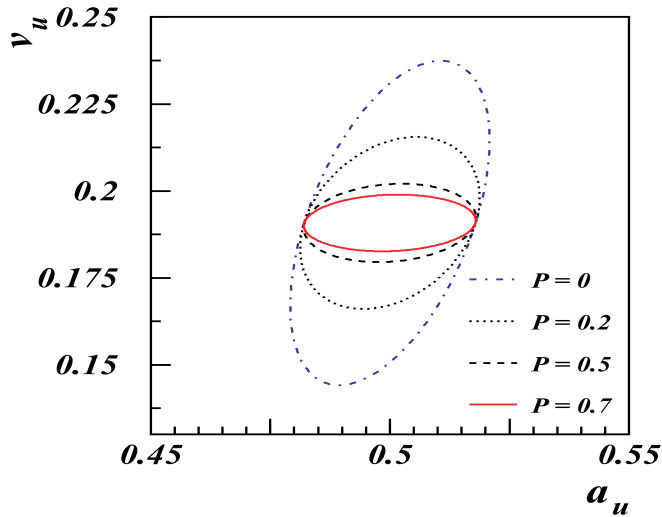}}
\caption{Sensitivity of the errors on the $u$ quark coupling to beam
polarisation, P (from \cite{light}).}
\label{fig:light}
\end{figure}
\noindent
As an example, the influence of the degree of polarisation, P, on the light 
quark coupling determination \cite{light} is presented in Fig. \ref{fig:light}.
The anticipated size of the error on the $u$ quark coupling decreases
significantly with increasing degree of the lepton beam polarisation. Possible
new physics can demonstrate itself by a deviation from the linear dependence of
the charged current cross section on the polarisation. Hera is also a potential
source of its ``own'' particle density functions. The PDFs measurement within
a single experiment over whole phase space offers a unique possibility of the
understanding, control and uniform treatment of systematic effects. Such
measurements require improved measurement and reconstruction of the kinematics
especially at high $x$ and $Q^2$. Since the tracks in such events are
collimated in the forward (proton) direction the detector performance in this
region has to be improved. Improvements of the vertexing and tracking will play
important role in the heavy quark studies after the upgrade. Presently, the
measurement of the $F^{c{\bar c}}_2$ \cite{f2c} suffers from the statistical
limitations of the data sample. It is clear that the precise tracking and
secondary vertex finding will improve the measurement. It will also help to
resolve the problem of the discrepancy between the data and the NLO
calculations of the $b$ quark production.\\
The ZEUS detector upgrade followed the physics programme. A new silicon 
micro-vertex detector (MVD) \cite{mvd} and the  forward, straw--tube tracker
(STT) \cite{stt} were installed. The MVD covers range of 10 -- 160 degrees of
the polar angle. It consists of three forward wheels and the barrel part
centered on the interaction point. This detector will be used for the precise 
tracking and the reconstruction of displaced vertices. The device will be used
for the study of the heavy flavor production and will improve the charged 
current physics. The second device, the STT, has four layers of the highly 
efficient, straw tube drift chambers. It extends the tracking range by more
than a unit in pseudorapidity  and covers the range of  5 -- 25 degrees of the
polar angle. Good tracking and vertexing abilities will help in the electroweak
and exotics studies with high $x$ and $Q^2$ events. It will be complementary to
the MVD for the heavy flavor studies. Also the vector meson and the QCD studies
will profit from the extension of the $W$ range.\\
Presently the ZEUS detector is tested and waiting for the data taking period.
\section{Acknowledgements}
I thank the organisers for the invitation and for the stimulating and enjoyable
atmosphere they provided. I am grateful to M. Derrick, M. Kuze and E. Tassi 
for discussions. The DESY Directorate financial support is kindly acknowledged.

%% file: quark02.bbl
\begin{thebibliography}{999}
\bibitem{zeusdet} The ZEUS Detector Status Report (unpubl.), U. Holm(ed), 
DESY, 1993.
\bibitem{tassi} ZEUS Coll., S. Chekanov et al., DESY-02-105.
\bibitem{cteq} J. Pumplin et al., Preprint hep-ph/0201195, 2002.
\bibitem{mrst} A. D. Martin et al., Preprint hep-ph/0110215, 2002.
\bibitem{bpcdata} ZEUS Coll., J. Breitweg et al., Eur. Phys. J. {\bf C~7}, 609
(1999),\\
ZEUS Coll., J. Breitweg et al., Phys. Lett. {\bf B~487}, 53 (2000).
\bibitem{moritz} ZEUS Coll., S. Chekanov et al., DESY-02-113,\\
ZEUS Coll., presented by M. Moritz at DIS 2002,
Cracow.
\bibitem{sjors} ZEUS Coll., S. Chekanov et al., Phys. Lett. {\bf B~539}, 197 (2002).
\bibitem{pdg} Particle Data Group, D. E. Groom et al., Eur. Phys. J. {\bf C~15} 1 (2000).
\bibitem{posmw} ZEUS Coll., J. Breitweg et al., Eur. Phys. J. {\bf C~12}, 411 (200).
\bibitem{h1wm} H1 Collab., C. Adloff et al., Eur. Phys. J. {\bf C~13}, 609 (2000).
\bibitem{dijetpap} ZEUS Coll., S. Chekanov et al., Eur. Phys. J {\bf C~23}, 13 (2002).
\bibitem{h1dj} H1 Collab., C. Adloff et al., DESY-01-225 and hep-ex/0201006.
\bibitem{xobs} ZEUS Coll., M. Derrick et al., Phys. Lett. {\bf B~348}, 665 
(1995).
\bibitem{frixione} S. Frixione, Z. Kunszt and A. Signer, Nucl. Phys. {\bf B~467}, 399 (1996),\\ 
S. Frixione, Nucl. Phys. {\bf B~507}, 295 (1997),\\
S. Frixione and G. Ridolfi, Nucl. Phys. {\bf B~507}, 315 (1997).
\bibitem{grv} M. Gl\"uck, E. Reya and A. Vogt, Phys. Rev. {\bf D~45}, 3986 
(1992),
Phys. Rev. {\bf D~46}, 1973 (1992).
\bibitem{apg} P. Aurenche, J. Guillet and M. Fontannaz, Z. Phys. {\bf C~64},
621 (1994).
\bibitem{cteq5m1} H. L. Lai et at., Phys. Rev. {\bf D~55}, 1280 (1997).
\bibitem{herwig} G. Marchesini et al., Comp. Phys. Comm., {\bf 67}, 465 (1992).
\bibitem{pythia} T. Sj\"ostrand, Comp. Phys. Comm., {\bf 82}, 74 (1994).
\bibitem{sigtot} ZEUS Coll., S. Chekanov et al., Nucl. Phys. {\bf B~627}, 3 (2002).
\bibitem{h1sigtot} H1 Coll., S. Aid et al., Phys. Lett. {\bf B~299}, 374 (1993), Z. Phys. {\bf C~69}, 27 (1995).
\bibitem{dl} A. Donnachie and P. V. Landshoff, Phys. Lett. {\bf B~437}, 408 (1998).
\bibitem{proceed} Future Physics at HERA, Proc. of the Workshop 1995/96, eds. 
G. Ingelmann, A. De Roeck, R. Klanner.
\bibitem{light} R. J. Cashmore et al., in Future Physics at HERA, Proc. of the
Workshop 1995/96, eds. G. Ingelmann, A. De Roeck, R. Klanner, p. 163.
\bibitem{f2c} ZEUS Coll., J. Breitweg et al., Eur. Phys. J. {\bf C~12}, 35 (2000).
\bibitem{mvd} E. Koffeman, NIM {\bf A~473}, 26 (2001).
\bibitem{stt} ZEUS Coll., A Straw-Tube Tacker for ZEUS, Jun. 1998 (unpubl.). 
\end{thebibliography}
